\documentclass[%
reprint,
 amsmath,amssymb,
 aps, 
 prl,
]{revtex4-2}

\usepackage{graphicx}% Include figure files
\usepackage{dcolumn}% Align table columns on decimal point
\usepackage{bm}% bold math
\usepackage{hyperref}% add hypertext capabilities
\usepackage[mathlines]{lineno}% Enable numbering of text and display math
\usepackage[dvipsnames]{xcolor}
\usepackage{mathptmx}
\usepackage{graphicx} 
\usepackage{float}
\usepackage{array}
\usepackage[inline]{trackchanges}

\addeditor{LK}
\addeditor{RA}
\addeditor{LC}

\hypersetup{
    colorlinks=true,
    linkcolor=blue,
    filecolor=magenta,      
    urlcolor=cyan,
}

\newcommand{\var}[1]{{#1}}

\begin{document}

\title{Using Thermal Crowding to Direct Pattern Formation on the Nanoscale}

\date{\today}% It is always \today, today,

\author{Ryan Allaire}
\affiliation{%
Department of Mathematical Sciences, United States Military Academy, West Point, NY, 10996
}   

\author{Linda J. Cummings, Lou Kondic}
%\email{kondic@njit.edu}
\affiliation{%
 Department of Mathematical Sciences, New Jersey Institute of Technology, Newark, New Jersey 07102, USA
}

\begin{abstract}
Metal films and other geometries of nanoscale thickness, when exposed to laser irradiation, melt and evolve as fluids as long as their temperature is sufficiently high.  This evolution often leads to pattern formation, which may be influenced strongly by material parameters that are temperature dependent.  In addition, the laser heat absorption itself depends on the time-dependent metal thickness.  Self-consistent modeling of evolving metal films shows that, by controlling the amount and geometry of deposited metal, one could control the instability development. In particular, depositing additional metal leads to elevated temperatures through the `thermal crowding' effect, which strongly influences the metal film evolution.  This influence may proceed via disjoint metal geometries, by heat diffusion through the underlying substrate. Fully self-consistent modeling focusing on the dominant effects, as well as  accurate time-dependent simulations, allow us to describe the main features of thermal crowding and provide a route to control fluid instabilities and pattern formation on the nanoscale.  
 \end{abstract}

\maketitle

\paragraph{Introduction}
Self- and directed-assembly has been the topic of numerous recent works, in particular for nanoscale fluid-based systems~\cite{lizmarzan_2010_acsnano,busnaina_2022_acsnano}.  While in many cases self-assembly can be used to produce patterns of interest, often directed assembly is needed to achieve desired outcomes such as, for example, an ordered array of nanoparticles. Systems involving metal films, filaments, and other geometries of nanoscale thickness are of particular current interest due to the large number of applications involving plasmonics, of relevance to solar cells, catalysis, and biomedical applications, etc., as reviewed by many authors~\cite{atwater_natmat10,wang2011,chaudhuri2011,Makarov2016,Hughes2017,ruffino_nano19}.  Such metal geometries are commonly exposed to short-duration (tens of nanoseconds) laser pulses to bring the material above the melting point. While molten, metals evolve as (to a first approximation, Newtonian) fluids, but with strongly temperature-dependent material properties, and are subject to fluid-dynamical instabilities.   Such instabilities, which evolve on a time scale comparable to that of the applied laser pulses, may lead to the formation of drops (becoming particles upon solidification). The size and placement of such particles are of crucial importance in applications, and it is important to be able to control them.

Directed assembly has been explored in the past using elaborately designed initial metal geometry, obtained for example by lithographically imposing sinusoidal perturbations to produce a desired outcome~\cite{fowlkes_nano11}. This method, while ingenious, may not be practical due to the need for costly lithographic-based modification of the initial metal  shape.
In this Letter we show that pattern formation can be controlled indirectly, via thermal transport through the supporting substrate.  We call this approach to directed-assembly `thermal crowding', since, as we will show, adjacent (though disconnected) metal geometries experience each other through thermal contact.
Therefore, simply by modifying the initial size and placement of simple metal structures such as filaments, one can direct the evolution and obtain patterns of desired properties.

\paragraph{Model} Modeling nanoscale metal films and other geometries exposed to laser irradiation is challenging since multiple physical effects must be included, particularly regarding the coupling of thermal effects due to the laser heating (including phase change) with fluid dynamics.  Our previous work in this area~\cite{arfm,allaire_jfm_2021,allaire_prf_2022}, which extended other research efforts, see, e.g.~\cite{trice_prb07,atena09,Saeki2013,shklyaev12}, constitutes a self-consistent and asymptotically-accurate framework for this complex problem.  The resulting set of equations, governing the metal film evolution while molten and the thermal transport, is discretized and solved in a GPU-based computing environment, using our open-source, publicly available code~\cite{GADIT_Thermal}.  Our particular focus is on evolving metal filaments of various sizes; such simple geometry illustrates the importance of careful coupling of the fluid dynamics with thermal transport and provides a straightforward proof-of-principle of using the initial configurations of simple metal shapes to control the final droplet (nanoparticle) size and number, without the additional complexities that could be anticipated for more elaborate initial geometries.  In what follows, we outline the main features of the theoretical and computational methods that we use and direct the interested reader to the supplementary material~\cite{sup_mat} and our earlier works~\cite{allaire_jfm_2021,allaire_prf_2022} for the details, including, in particular, the careful asymptotic expansion of the governing equations that leads to the formulation presented here.

\begin{figure}[t]
    \centering
    \includegraphics[width=0.5\textwidth]{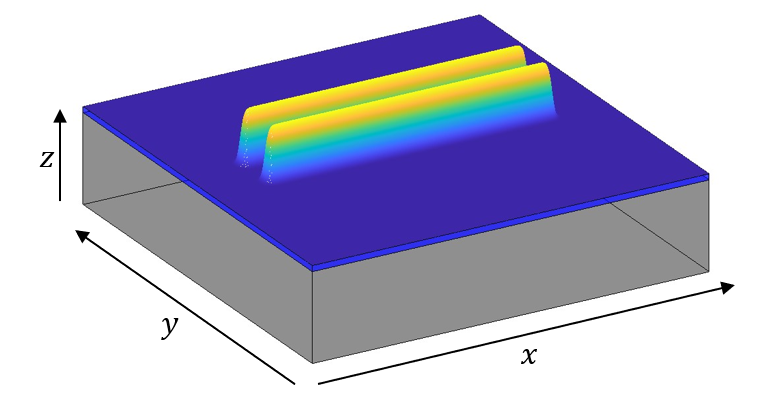}
    \caption{Schematic of two neighboring filaments (both indicated by the color gradient, with yellow representing maximum height) and thinner equilibrium film region (dark {\color{blue} blue}). The gray region represents the underlying SiO$_2$ substrate. In physical experiments~\cite{McKeown2012}, the in-plane dimensions are measured in hundreds of nm, typical metal thickness is 10-20 nm, and the substrate (membrane) thickness is on the order of 100 nm.}
    \label{fig:schematic1}
\end{figure}

Consider a free surface metal filament of characteristic nanoscale thickness, $H=10$ nm, which we will use as a length scale. Suppose further that the metal is initially solid, exposed to air above, and in contact below with a thermally conductive solid SiO$_2$ substrate of depth $H_{\rm s}$, see Fig.~\ref{fig:schematic1}. In order to develop a model, we make a number of simplifying assumptions, which we outline after presenting the model. Following~\cite{allaire_jfm_2021}, we choose the velocity scale  $U=\gamma_{\rm f}/(3 \mu_{\rm f})$ (where $\gamma_{\rm f}$ and $\mu_{\rm f}$ are surface tension and viscosity at melting temperature, $T_{\rm melt}$), leading to the time scale, $H/U$. Subsequently, we choose $T_{\rm melt}$, $\mu_{\rm f} U/H$ and $\gamma_{\rm f}$ as the temperature, pressure, and surface tension scales, respectively. We take the dimensionless domain length/width (in the $x$/$y$ directions) to be $P$.

Once in the liquid state, we treat the metal as an incompressible Newtonian fluid. The most complete model consists of the Navier-Stokes equations for the molten metal film, coupled with heat equations for the metal and substrate, plus appropriate boundary conditions and constitutive equations for the temperature-dependent material parameters. We assume that the viscosity depends on temperature, but that the surface tension, density, heat capacity and thermal conductivity of the material are fixed at their melting temperature values (see Sec. II of \cite{sup_mat} for a justification of neglecting the temperature dependence of surface tension, and prior work~\cite{allaire_prf_2022} for discussion of the temperature dependence of thermal conductivity). Note in particular that the model allows for both spatial and temporal dependence of viscosity through the dependence of this material parameter on the temperature field.  We use $T_{\rm f}$, $T_{\rm s}$ to denote the film and substrate temperatures, respectively and define $T_{\rm avg}(t)$, to be the time-dependent average filament temperature (see Sec. II of \cite{sup_mat} for a complete definition). For the remainder of the text, we omit the argument of $T_{\rm avg}(t)$ with the understanding that it is time-dependent. 

In recent work~\cite{allaire_jfm_2021, allaire_prf_2022} we proposed a simplified model for this setup based on a long-wave formulation for a thin metal geometry. The metal film thickness ($z=h(x,y,t)$ where $x$ and $y$ are the in-plane coordinates) evolution is governed by the following 4th order nonlinear partial differential equation, 
\vspace{-0.1in}
\begin{align}
    \partial_t h + \nabla_2 \cdot \left[ \frac{1}{\mu(T_{\rm f})} h^3  \nabla_2 \left( \nabla_2^2 h + \Pi(h) \right)  \right]  = 0, \label{eq:thin_film}  
    \vspace{-0.1in}
\end{align}
where $\nabla_2=(\partial_x,\partial_y,0)$. Following the time derivative term in Eq.~\eqref{eq:thin_film} are the capillary and disjoining pressure terms, respectively.  Here, $\mu(T_{\rm f})$ represents temperature-dependent viscosity (scaled by $\mu_{\rm f}$), modeled via an Arrhenius-type relationship, 
% We model the temperature dependence of viscosity $\mu(T_f)$ with an Arrhenius-type relationship,
\vspace{-0.1in}
\begin{align}
    \mu (T_{\rm f}) = S\left(T_{\rm f}\right)\exp\left( \mathcal{E}\left( \frac{1}{T_{\rm f}}-1 \right) \right), \label{thicksub_viscosity_eq} %\\
    %S\left( T_{\rm f}\right)=\frac{1}{2} \left( \tanh \left( (T_{\rm %f}-1) + \delta_T \right) + 1 \right) \label{eq:sigmoid}
\end{align}
%\vspace{-0.1in}
where $\mathcal{E}= E/(R T_{\rm melt})$ is related to the activation energy $E$ ($R$ is the universal gas constant)~ \cite{metals_ref_book_2004}. %Here, we follow the approach developed in~\cite{seric_pof18} in utilizing Eq.~(\ref{thicksub_viscosity_eq}).  
This form also allows for melting/solidification control; the Sigmoid function 
$S\left( T_{\rm f}\right)= 2 \left( \tanh \left( (T_{\rm f}-1) + \delta_T \right) + 1 \right)^{-1}$ approximates the phase transition ($\delta_T$ is set to $5 {\rm K}/T_{\rm melt}$). 

For metal films of nanoscale thickness, instability due to destabilizing metal-substrate interaction is an important effect since the range of the interaction potential is comparable to the film thickness~\cite{Israelachvili}.  This is modeled via a disjoining pressure of the form  $\var{\Pi}(\var{h})= \Omega [ (h_*/h)^n - (h_*/h)^m ]$, with equilibrium film thickness ${h}_*$, constant $\Omega$ (related to the Hamaker constant $A_{\rm H}$ by $\Omega= {A_{\rm H}/{(6\pi \gamma_{\rm f} h_*^3 H^2)}}$), and exponents $n>m>1$~\cite{DK2016}. Such a disjoining pressure term ensures that the film height nowhere goes to zero, but instead approaches a minimum value comparable to $h_*$ as dewetting proceeds (we typically use $h_* = 0.1$, corresponding to 1 nm; while this value is larger than that expected in experiments~\cite{Gonzalez2013}, this choice avoids the numerically more expensive simulations that are required for smaller values of $h_*$).
Interfacial potentials for liquid metals are undoubtedly more complex than specified here, however, based on our earlier work and extensive comparison to experiment (see in particular~\cite{Gonzalez2013}), we expect that the present form is sufficient for our purposes; further details on disjoining pressure models in this context are given in a recent review~\cite{arfm}.

The heat flow model we developed recently~\cite{allaire_jfm_2021} exploits further the long-wave approximation. The metal has much higher thermal conductivity than the substrate, and the temperature variation across the film (in the short, $z$-direction) may be shown to be weak. We retain those assumptions here, but different to our earlier work (and justified below), we assume that in-plane heat conduction in the substrate may be relevant and that heat is lost from the system only through the lateral substrate boundaries. This leads to the following system governing temperature,
\begin{align}
h \mbox{Pe}_{\rm f} \partial_t T_{\rm f} &= \nabla_2 \cdot \left( h \nabla_2 T_{\rm f} \right) - \mathcal{K} \partial_z T_{\rm s}\vert_{z=0} + h \overline{Q},
\label{eq:tm}
\\
\mbox{Pe}_{\rm s}\partial_t T_{\rm s} & =  \nabla_2^2 T_{\rm s} + \partial_z^2 T_{\rm s},
\label{eq:ts}
\end{align}
for $z \in \left(0,h\right)$ and $z \in \left(-H_s,0\right)$, respectively.  The boundary conditions (BCs) include continuity of temperature at $z=0$ and appropriate BCs at the domain boundaries: 
$T_{\rm f} = T_{\rm s}$ at $z = 0$, 
$\partial_z T_{\rm s} = 0$ at $z=-H_s$, 
$\partial_x T_{\rm f} =0$ at $x=0, P$, 
$\partial_y T_{\rm f} =0$ at $y=0, P$, and 
$T_{\rm s} = T_{\rm a}$ (ambient temperature) at $x = 0, P$, $y=0,P$.
The parameters, defined by
 \begin{align}
\mbox{Pe}_{\rm f} &= \frac{\left( \rho c \right)_{\rm f} U H }{k_{\rm f} }, \qquad \mbox{Pe}_{\rm s} = \frac{\left( \rho c \right)_{\rm s} U H }{k_{\rm s} }, \qquad \mathcal{K}=\frac{k_{\rm s}}{k_{\rm f}},  \nonumber 
 \end{align}
represent the film and substrate Peclet numbers, and thermal conductivity ratio, respectively. Equation \eqref{eq:tm} describes the leading order fluid temperature and the terms on the right-hand side represent the in-plane diffusion, heat loss due to the substrate, and heat generated due to the laser, averaged over the metal thickness,
\begin{align}
\overline{Q}= \frac{A(h)}{h} \int_{0}^{h} F(t) \left[1-R(h) \right] \exp \left[ -\alpha_{\rm f} \left(h-z \right) \right]dz .
\end{align}
Here, $\alpha_{\rm f}^{-1}$ is the absorption length for laser radiation in the metal film and $F(t)$ captures the temporal shape of the laser pulse, taken to be Gaussian centered at specified time $t_{\rm p}$ and of prescribed width $\sigma$ (corresponding to tens of nanoseconds). The term $A(h)$ is a smooth approximation of $\mathcal{U}(h-h_*)$, the unit step function centered at $h_*$, which turns off absorption as $h\rightarrow h_*$.  In-plane diffusion is similarly turned off as $h\rightarrow h_*$, which ensures that the filaments alone absorb energy and transfer heat to other filaments only via the substrate. In general, the reflectivity of the film on a transparent substrate, $R(h)$, is found by solving Maxwell's equations~\cite{heavens55}, but the resultant form is cumbersome; following earlier work~\cite{seric_pof18,trice_prb07} we approximate it by  $R(h)=r_0 \left(1-\exp \left(-\alpha_{\rm r} \var{h} \right) \right)$, where $r_0$ and $\alpha_{\rm r}$ are parameters, chosen to ensure good agreement with the exact solution.

Equation \eqref{eq:ts} describes the substrate heat conduction.  
The BCs on Eqs.~(\ref{eq:tm}) and (\ref{eq:ts}) impose  
no heat loss at the bottom of the substrate (motivated by the experiments on thin substrates (membranes) discussed below), as well as insulating conditions for the metal at its lateral boundaries. We impose room temperature at the lateral boundaries of the substrate, leading to the only heat loss mechanism.  The no-heat loss BC at the bottom of the substrate motivates the inclusion of in-plane thermal flow in Eq.~(\ref{eq:ts}) since there is no significant heat flow in the $z$ direction.

In deriving our model, we made the following additional assumptions not yet discussed: 
 (i) inertial effects are negligible (discussed and justified elsewhere~\cite{gonzalez_jfm16}); (ii) phase change (melting, solidification) is fast, and the associated energy gain/loss can be ignored (see e.g.~\cite{seric_pof18,trice_prb07}); (iii) heat is lost from the metal only through the substrate with no radiative losses~\cite{allaire_prf_2022}; (iv) the metal does not evaporate.  We also reiterate that the equilibrium layer of thickness $h_*$ plays no role in thermal transport; we remove heat diffusion (the first term on the right-hand side of Eq.~(\ref{eq:tm})) through this layer to reduce its role to modeling fluid-dynamical aspects of the problem only.  

\paragraph{Results}
We focus on setups relevant to recent experiments~\cite{diez_langmuir_2021} involving nanoparticle formation on so-called `membranes'. Membranes are essentially very thin solid substrates with overlaid nanoscale metal patterns, obtained by combining lithographic techniques with chemical etching of the underlying silicon~\cite{diez_langmuir_2021}.  An important motivation for using membranes is that one is able to observe not only the final outcome of the experiments but also the time evolution since membranes are optically transparent and allow for the use of dynamic transmission electron microscopy (DTEM), which provides unique nanosecond temporal and nanometer spatial resolution~\cite{diez_langmuir_2021}.  From the modeling perspective, the key point is that the Biot number is typically very small so the bottom boundary of the substrate is essentially insulating. Therefore, the membrane setup allows for precise control of heat flow. We note that, while carrying out experiments on (thick) SiO$_2$ wafers requires more energy, the model that we have discussed here describes such experiments accurately as well, see~\cite{allaire_prf_2022} for details. 

In our simulations, the initial condition is a metal filament, possibly surrounded by other filaments at a specified distance apart, with the exact geometry specified in Sec.~I of \cite{sup_mat}. Initially, both metal and substrate are at room temperature. Then, the laser pulse is applied, the metal temperature increases (discussed more precisely in what follows), and when the filament temperature rises above the melting point, it starts evolving as a Newtonian fluid. When the laser energy begins to decrease, the filament temperature decreases as well, and once it drops below the melting temperature, the evolution stops. 
The simulations are carried out using our in-house GPU-based code~\cite{GADIT_Thermal}.  The computational method itself is based on finite difference spatial discretization, combined with Crank-Nicolson temporal evolution within an ADI (alternate direction implicit) framework; see~\cite{allaire_prf_2022} for details. The parameters used in all simulations are provided in Table 1 of \cite{sup_mat}.  

%% Proof of principle plot
\begin{figure}[t]
    \centering
    \includegraphics[width=0.5\textwidth]{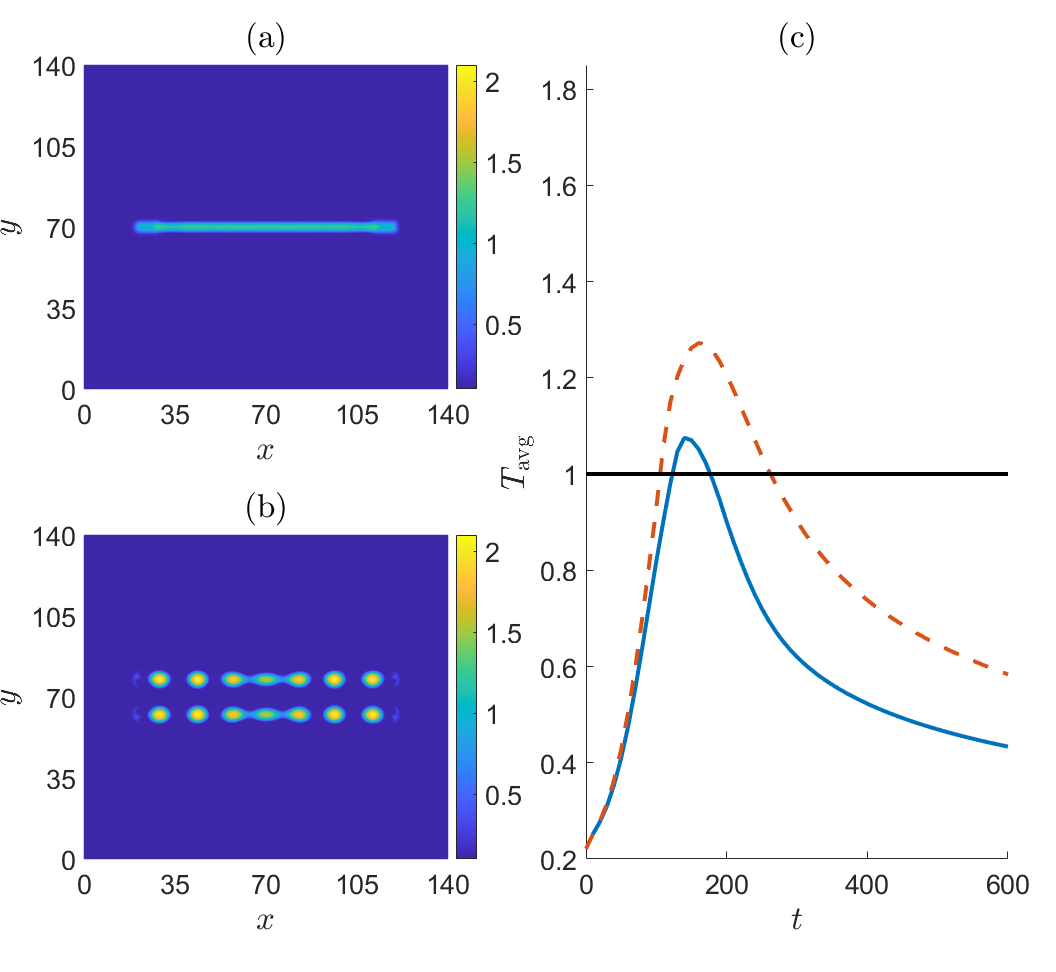}
    \caption{Final configuration of (a) a single metal filament and (b) two metal filaments placed at a distance of $D=15$ apart, exposed to the same laser pulse.
    %For both (a) and (b), the dark blue region indicates the equilibrium film region, which is not heated. 
    Frame (c) shows the maximum temperatures for the configurations in (a) ({\color{blue} blue} solid line) and (b) ({\color{red} red} dashed line), along with the melting line (black). See also Animation 1~\cite{sup_mat_anim} for fluid evolution and temperature fields.
    }
    \label{fig:pillar_existence}
\end{figure}

Figure \ref{fig:pillar_existence}(a~-~b) shows the results of two sets of simulations, whose only difference is the number of filaments present. Still, the final outcome is very different.  While in both cases the filament melts, no significant instability development has occurred in (a) prior to resolidification, whereas in (b) the evolution is much more advanced, exhibiting a pearling type of instability~\cite{dk_pof07,dgk_pof09}.
Figure \ref{fig:pillar_existence}(c) shows why: the average temperature is significantly higher in the case of two filaments, and furthermore, the metal temperature in this case remains above melting for a longer time. Due to the Arrhenius-type dependence of viscosity on temperature, the evolution is also faster when multiple filaments are present.  Animation 1~\cite{sup_mat_anim} shows the time- and space-dependence of the film and temperature evolution; note that filaments cool faster towards their ends.
The animation also emphasizes the significantly higher temperature attained for two filaments compared to one, leading to the faster evolution and breakup noted above. We remark that the details of the final outcome depend on the choice of parameters, including the filaments' volume and aspect ratio, see~Secs.~V and VI of~\cite{sup_mat}; the concept of thermal crowding is, however, always found to hold, with faster evolution and breakup for multiple filaments.

\begin{figure}[t]
    \centering
    \includegraphics[width=0.5\textwidth]{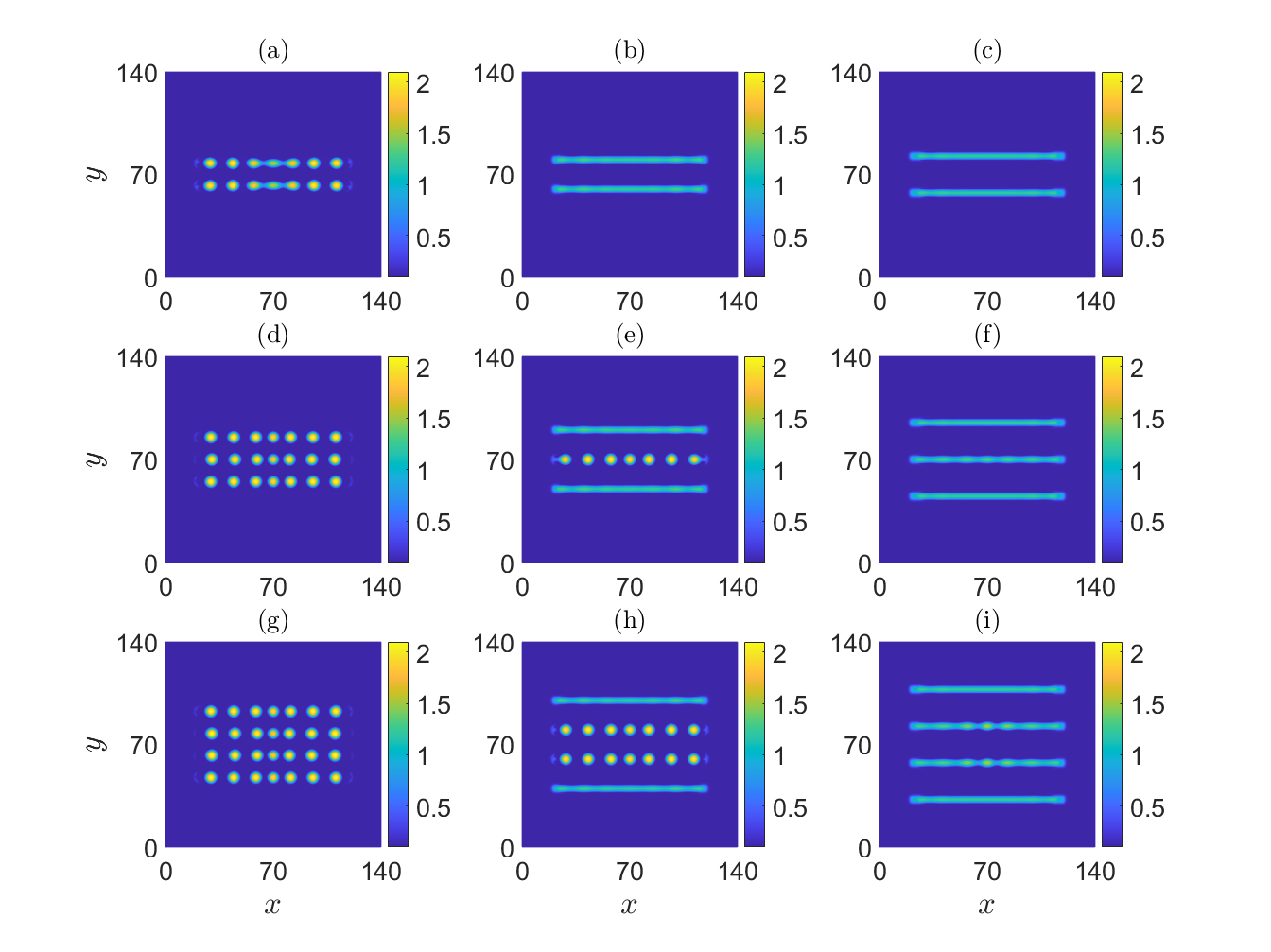}    
    \includegraphics[width=0.5\textwidth]{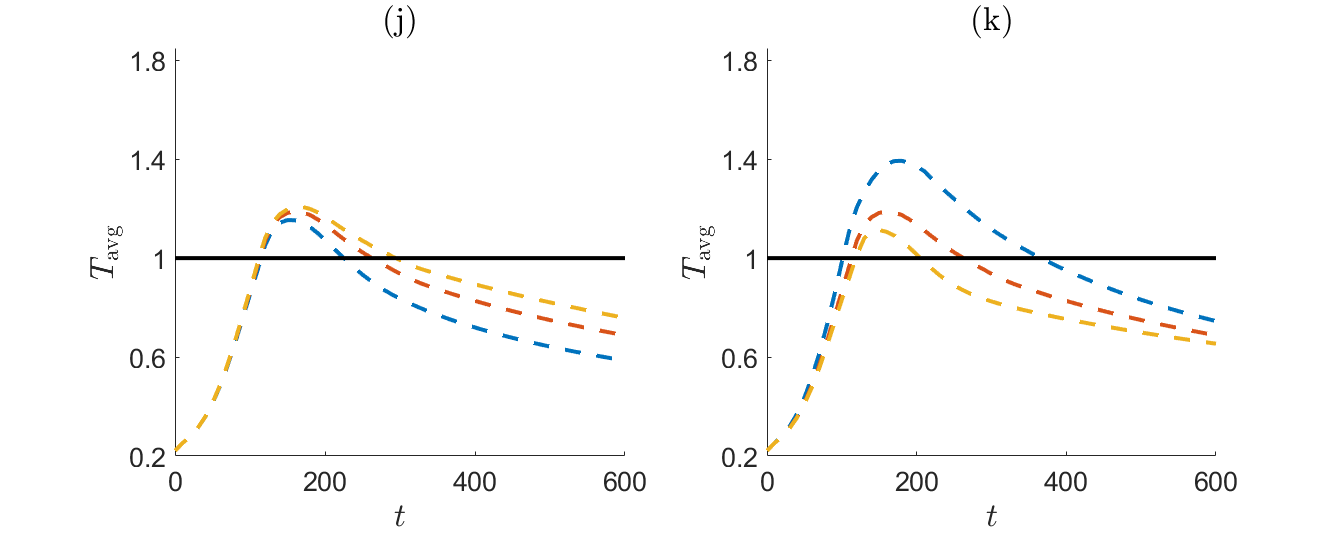}
    \caption{(a-i) Final configuration of two (row 1), three (row 2), and four filaments (row 3) at distances $D=15$ (column 1), $D=20$ (column 2), $D=25$ (column 3). (j) Time evolution of the average film temperature for (b) {\color{blue} blue} dashed line, (e) {\color{orange} orange} dashed line, and (h) {\color{Dandelion} yellow} dashed line. (k) Time evolution of the average film temperature for (d) {\color{blue} blue} dashed line, (e) {\color{orange} orange} dashed line and (f) {\color{Dandelion} yellow} dashed line.  The melting temperature, $T_{\rm melt}$, is shown by the solid black line. See also Animations 2 and 3 at~\cite{sup_mat_anim}.    
       }
    \label{fig:pillar_size}
\end{figure}

Figure \ref{fig:pillar_size} shows the effect that the number of filaments and their respective spacing has on the collective heating and dewetting, illustrating that thermal crowding can strongly influence filament evolution.  The comparison of different rows shows that an increase in the number of filaments strongly influences their temperatures and, through the temperature-dependent viscosity, the resultant dewetting. For example, in Fig.~\ref{fig:pillar_size}(b) the filaments melt but not sufficiently even to dewet partially, whereas the filaments in Fig.~\ref{fig:pillar_size}(e, h) both show full dewetting of the interior filaments. Interestingly, the filaments that are farthest from the center only begin undulation growth before they freeze in place because they only receive one-sided diffusive heating from the other filaments. The interior filaments of Fig.~\ref{fig:pillar_size}(e, h) then spend more time in the liquid phase, are hotter, of lower viscosity, and therefore evolve faster than the outermost filaments; see also Animation 2~\cite{sup_mat_anim}. This animation also illustrates the relevance of both the number of filaments and their spacing; in the particular case considered here, the distance between filaments is more important since the configuration in Fig.~\ref{fig:pillar_size}(d) (three filaments at spacing $D=15$) remains hot longer than that in Fig.~\ref{fig:pillar_size}(h) (four filaments at spacing $D=20$), consistent with the information on average temperatures provided by Fig.~\ref{fig:pillar_size}(j, k). 

In principle, there are two possible reasons for the unbalanced heating leading to a breakup of internal filaments only: (i) loss of heat through the domain boundaries, and also (ii) decreased heating of the external filaments by the internal ones. Insight regarding which of these two mechanisms is most relevant can be reached by carrying out simulations on larger domains. These additional results, see~Sec.~IV of \cite{sup_mat}, show that the influence of the heat loss through domain boundaries is minimal, and therefore, one-sided heating governs the evolution and breakup.

Figure~\ref{fig:pillar_size}(j) shows that increasing the number of filaments, while keeping their spacing fixed, has a fairly modest effect on the maximum metal temperature, but a larger number of filaments collectively retains heat for longer, leading to increased liquid lifetimes and more complete dewetting in the cases with three and four filaments. 
Figure~\ref{fig:pillar_size}(k) shows that increasing the inter-filament distance, $D$, leads to lower filament temperatures. In particular, placing three filaments at a distance $D=15$ (Fig.~\ref{fig:pillar_size}(d)) is sufficient to collectively melt and fully evolve all three into nanoparticles, as opposed to placing them at distance $D=25$ (Fig.~\ref{fig:pillar_size}(f)), which (although the filaments melt) is insufficient to produce any nanoparticle formation. Additional information regarding spatial temperature distribution is available in Animation 3,
which shows that the temperature maxima occur at the center of the domain, with lower temperatures towards the periphery, consistent with, e.g., the results shown in Fig.~\ref{fig:pillar_size}(i) where for the central filaments, we observe better-developed undulations towards the filament centers.

\begin{figure}[t]
    \centering
    \includegraphics[width=0.5\textwidth]{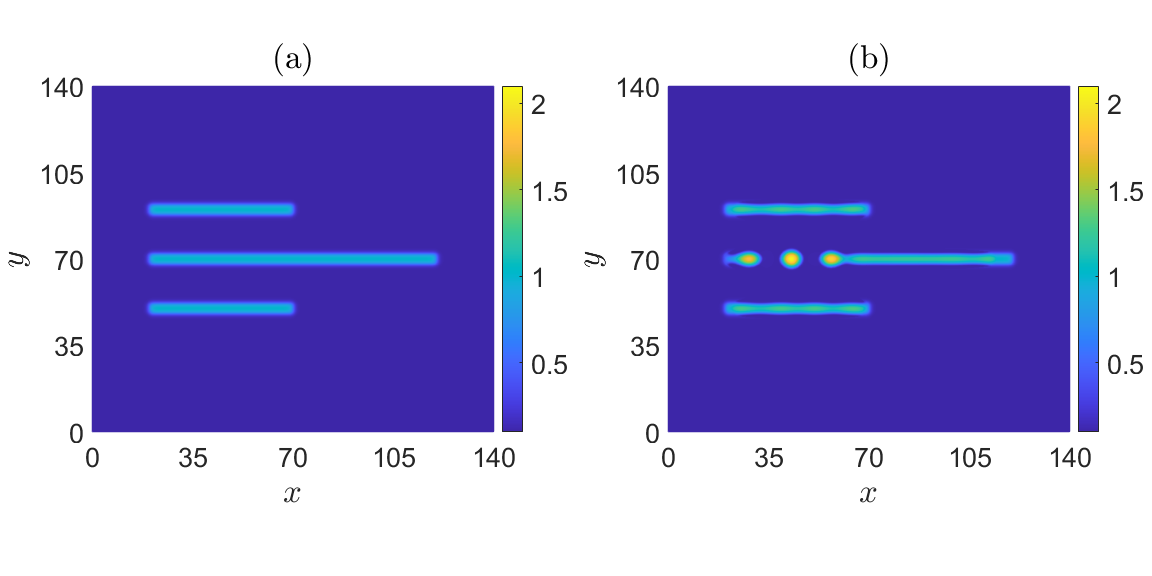}
    \caption{Initial (a) and final (b)configuration for asymmetric setup with a filament of length $100$ surrounded by two filaments of length $50$ placed at a distance $D=20$.
    See also Animation 4 at~\cite{sup_mat_anim}. }
    \label{fig:left}
\end{figure}

Figure~\ref{fig:left} illustrates that filament instability can be initiated asymmetrically.  Here, we place short filaments on both sides of the long filament that was shown in Fig.~\ref{fig:pillar_existence}(a), which alone does not break up. 
The presence of the additional short filaments is, however, sufficient to increase the temperature of the left side of the long filament and induce breakup.  The short filaments themselves melt; however, asymmetric one-sided heating is insufficient for their complete breakup. Additionally, Animation 4~\cite{sup_mat_anim} shows that the leftmost droplet solidifies prior to fully dewetting, illustrating the nontrivial competition between the edge retraction and the resolidification. Additional examples of more elaborate filament configurations are given in Sec. VII of~\cite{sup_mat}.

\paragraph{Conclusions}
We have illustrated a simple but powerful method that allows for the coupling of fluid dynamics and heat transport for metal filaments deposited on thermally conductive substrates.  Our approach is fully self-consistent, with fluid dynamics influencing and being influenced by the heat flow.  This coupling occurs through temperature dependence of metal viscosity, whose spatial variation influences the fluid thickness and, in turn, affects the amount of heating absorbed. In the present work, we apply this method to metal filaments; however, the model can be used for any material and any initial material geometry exposed to volumetric heating.  In the context of metals, our results open the door to various directed- and self-assembly approaches since it is now possible to control the dynamics simply by specifying the initial material distribution.

\paragraph{Acknowledgement}
This research was supported by NSF DMS-1815613, an NJIT seed funding grant (2022), and a USMA Dean's Faculty Research Fund.  

\bibliography{films}

\end{document}

% \maketitle

\begin{center} \large 
    Supplementary Material \\ \vspace{0.3in}    
        {\bf Using Thermal Crowding to Direct Pattern Formation on the Nanoscale} \\ \vspace{0.2in}       by \\ \vspace{0.2in}       Allaire, Cummings, Kondic
\end{center}

\section{Initial Condition}
\begin{figure}[b]
    \centering
    \includegraphics[width=0.75\textwidth]{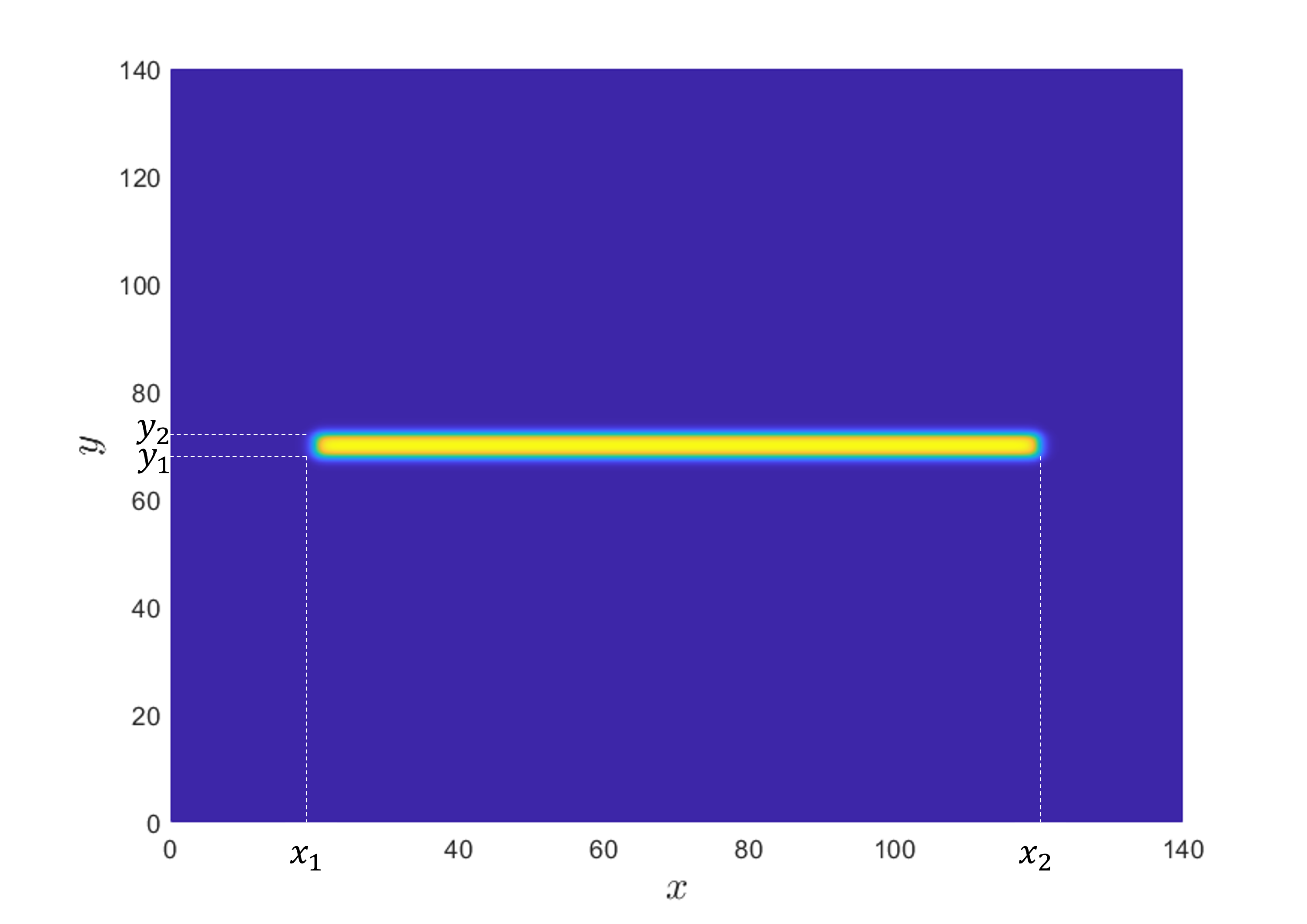}
    \caption{Filament geometry. Filament of length $x_2-x_1$, width $y_2-y_1$ (both marked with dotted white lines) and height of $h=1$. The {\color{blue} dark blue} region indicates the equilibrium film, where $h=h_*$ and the {\color{goldenyellow} yellow} region indicates the filament. Intermediate colors indicate the smooth transition region.}
    \label{fig:IC_schematic}
\end{figure}

The initial condition of a single filament is given by 
\begin{align*}
    h(x,y,0) = \mathcal{F}(x,y) =  h_* + A_0 \left( \tanh(x-x_1) \tanh(x_2 - x) + 1 \right) \left(  \tanh(y-y_1) \tanh(y_2 - y) + 1 \right), 
\end{align*}
where the parameters $x_1=20$ and $x_2=120$ are the lateral boundaries of the filament (making a filament of length  $L=100$ in dimensionless units) and $y_1=68$, $y_2=72$ are the transverse boundaries (setting the filament width of $4$ dimensionless units). The scaling factor $A_0$ determines the height of the filament, which remains at $1$, unless otherwise stated. Figure \ref{fig:IC_schematic} shows the initial condition of a single filament with $x_1,x_2, y_1$, and $y_2$ labeled on a domain of $P\times P = 140\times 140$. The dark blue region indicates the equilibrium film  whereas the yellow region represents the filament.  The specified values of $L$ and $P$ are used throughout except if specified differently. 

The multi-filament initial conditions are then created via a combination of single filaments, $h(x,y,0)=\sum \mathcal{F}_i(x,y)$, where each $\mathcal{F}_i$ is a shift of $\mathcal{F}$ in the $y-$direction. For filament configurations of varying length, the length of each $\mathcal{F}_i$ is controlled by modifying $x_1$ and $x_2$.

\section{Effect of Surface Tension}\label{sect:effect_of_surf_tens}
In this section we investigate the effect of varying surface tension with temperature on the dewetting of the filaments presented in the main text. When surface tension is allowed to vary with temperature the evolution of the film is governed by (see~\cite{allaire_prf_2022})
\begin{align}
    \partial_t h + \nabla_2 \cdot \left[ \frac{1}{\mu(T_{\rm f})} \left(  h^3  \nabla_2 \left( \Gamma \nabla_2^2 h + \Pi(h) \right) + h^2 {\rm Ma} \nabla_2 \left( \Delta T \right)  \right) \right]  = 0, \label{eq:thin_film_supp}  
\end{align}
where $\Gamma = 1+2{\rm Ma}(T_{\rm avg}-1)/3$ is the leading order surface tension (scaled by $\gamma_{\rm f}$) that depends linearly on the average filament temperature $T_{\rm avg}$. ${\rm Ma}=3\gamma_{\rm T} T_{\rm melt}/(2 \gamma_{\rm f})$ is the Marangoni number accounting for variation of surface tension due to temperature gradients at the metal surface, $\Delta T$. Also, $\gamma_T=(\gamma_{\rm f}/T_{\rm melt})\mathrm{d}\gamma/\mathrm{d}T_{\rm avg} \vert_{T_{\rm avg}=1}$ is the change in surface tension with temperature when the filament (on average) is at melting temperature, $T_{\rm avg}=1$. The average filament temperature is computed via

\begin{align}
T_{\rm avg}(t)=\frac{1}{P^2}\int_{0}^{P} \int_{0}^{P} \mathcal{U}(h-1.1h_*) T_{\rm f}(x,y,t)\:\mathrm{d} x \mathrm{d} y, \label{thicksub_T_bar_defn}
\end{align}
where $\mathcal{U}(h-1.1h_*)$ is the unit step function that ensures the average is taken only over the filament region (defined as regions of thickness $10\%$ above the equilibrium thickness, $h_*$). The two differences between Eq.~\eqref{eq:thin_film_supp} and Eq. (1) of the main text is the addition of variable $\Gamma$ and the Marangoni term, ${\rm Ma}$. 

\begin{figure}[t]
    \centering
    \includegraphics[width=0.6\textwidth]{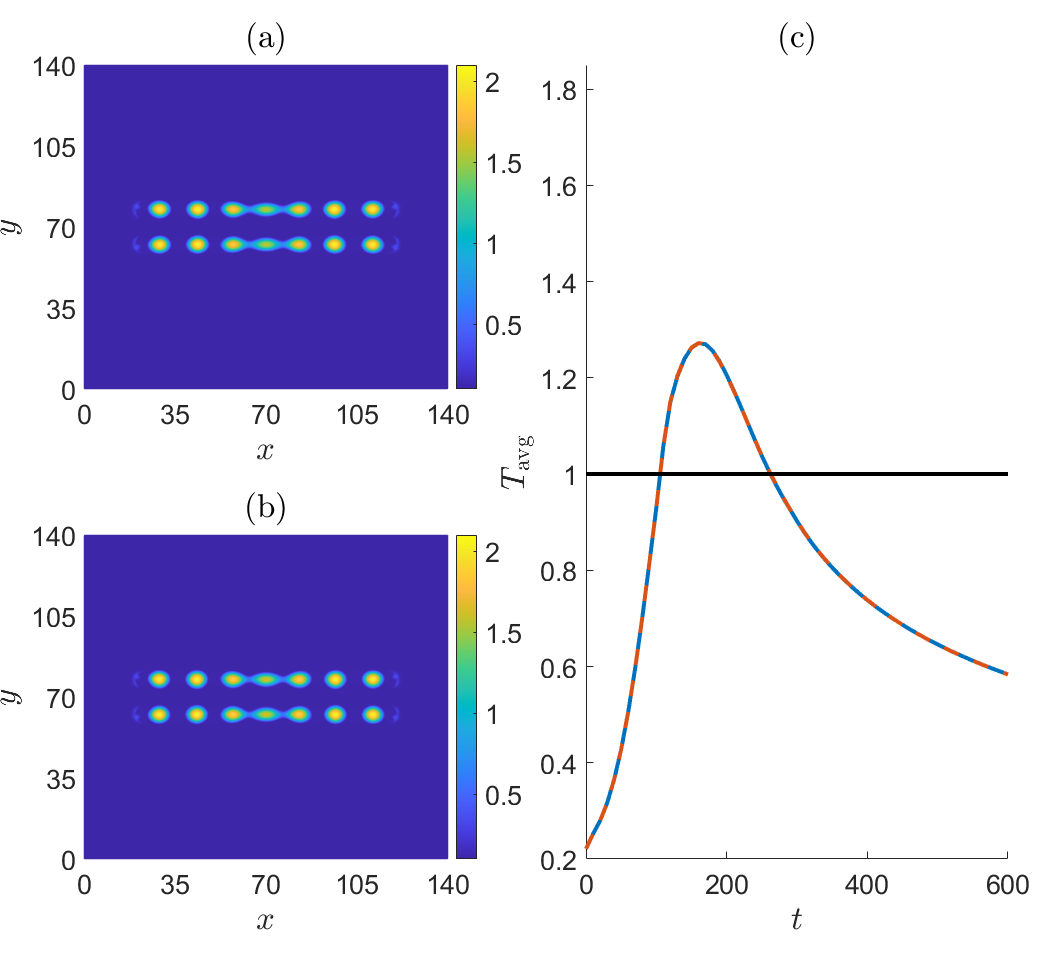}
    \caption{Influence of temperature-dependent surface tension.  Final configuration of two filaments, separated by a distance $D=15$ where (a) surface tension is constant (reproduced from Fig.~2(b) in the main text to simplify the comparison), and (b) temperature-dependence of surface tension is included through both the Marangoni effect and average temperature dependence, ${\rm Ma}\neq 0, \Gamma=\Gamma(T_{\rm avg})$. The average filament temperatures are given in (c) where the {\color{blue} blue} solid line corresponds to (a) and the {\color{red} red} dashed line corresponds to (b). The melting temperature is shown by the black solid line.}
    \label{fig:effect_of_marangoni}
\end{figure}

Figure \ref{fig:effect_of_marangoni} shows the influence of temperature-dependent surface tension, both through the Marangoni effect (gradients of surface tension with space through temperature gradients) and average filament temperature dependent leading order surface tension, on the evolution of two filaments placed at distance $D=15$ apart (this value of $D$ is used throughout if not specified differently). Part (a), which shows the evolution with constant surface tension (${\rm Ma}=0,~ \Gamma=1$), is indistinguishable from that of (b), which shows the filament evolution with temperature-dependent surface tension (${\rm Ma}\neq 0, \Gamma=\Gamma(T_{\rm avg})$). Part (c) shows the consistent average filament temperatures for both (a) ({\color{blue} blue} solid line) and (b) ({\color{red} red} dashed line). Combined with (a) and (b) this indicates that the variation of surface tension with temperature (both spatial and temporal variations) does not significantly alter the heating and evolution of the filament, and justifies keeping surface tension fixed at its melting temperature value in the main text, excluding both the Marangoni effect and fixing $\Gamma=1$ . Although the simultaneous effect of ${\rm Ma}$ and $\Gamma(T_{\rm avg})$ are shown here in (b), both effects were verified to produce analogous results to that shown in Fig.~\ref{fig:effect_of_marangoni} and are omitted for simplicity.

%% Parameter Section
\section{Parameters}
Here we provide the values of the parameters used in the simulations. Table \ref{table:ref_paras} shows the values for liquid ${\rm Cu}$ on top of a ${\rm SiO}_2$ substrate. 
\begin{table}[t]
\centering
{\renewcommand{\arraystretch}{0.75} % for the vertical padding
  \begin{tabular}{ | l | l | l | l |} \hline 
 \textbf{Parameter} & \textbf{Notation} & \textbf{Value} & \textbf{Unit}  \\ \hline 
  Viscosity at $T_{\rm melt}$ & $\mu_{\rm f}$ \cite{DK2016}  & $4.3 \times 10^{-3}$ & $\mathrm{Pa\, s}$  \\  
  Surface tension at $T_{\rm melt}$ & $\gamma_{\rm f}$ \cite{DK2016}   & 1.303 & $\mathrm{J\, m^{-2}}$ \\ 
  Length scale & $H$ & $10$ & $\mathrm{nm}$  \\   
  Time scale & $t_{\rm scl} = 3H\mu_{\rm f}/ \gamma_{\rm f}$ & $0.09$ & $\mathrm{ns}$ \\  
Melting Temperature & $T_{\rm melt}$ & $1358$ & $\mathrm{K}$ \\ 
    
  Film density & $\rho_{\rm f}$ \cite{DK2016}   & $8000$ & $\mathrm{kg \, m^{-3}}$ \cite{DK2016}  \\  
  SiO$_2$ density & $\rho_{\rm s}$ \cite{DK2016}    & $2200$ & $\mathrm{kg \, m^{-3}}$  \\ 
  Film specific heat capacity & $c_{\rm f}$ \cite{DK2016}   & $495$ & $\mathrm{J \, kg^{-1} \, K^{-1}}$ \\  
  SiO$_2$ specific heat capacity & $c_{\rm s}$ \cite{DK2016}  & $937$ & $\mathrm{J \, kg^{-1} \, K^{-1}}$ \\  
  Film heat conductivity & $k_{\rm f}$ \cite{DK2016}   & $340$ & $\mathrm{W \, m^{-1} \, K^{-1}}$  \\ 
    SiO$_2$ heat conductivity & $k_{\rm s}$ \cite{allaire_prf_2022}  & $1.4$ & $\mathrm{W \, m^{-1} \, K^{-1}}$\\  
  Film absorption length & $\alpha_{\rm f}^{-1} H$ \cite{DK2016}   & $11.09$ & $\mathrm{nm}$  \\ 
  Temp. Coeff. of Surf. Tens. & $ \gamma_{\rm T}$ \cite{DK2016}   & $-0.23 \times 10^{-3}$ & $\mathrm{J \, m^{-2} \, K^{-1}}$  \\  
    Hamaker constant & $A_H$ \cite{Gonzalez2013}  & $3.49\times 10^{-17}$ & $\mathrm{J}$  \\
  Reflective coefficient & $r_0$ \cite{DK2016}   & $0.3655$ & 1  \\ 
  Film reflective length & $\alpha_{\rm r}^{-1} H$ \cite{DK2016}   & $12.0$ & $\mathrm{nm}$ \\  
    Laser energy density & $E_0$  & $6800$ & $\mathrm{J \, m^{-2}}$  \\ 
    Gaussian pulse peak time & $t_{\rm p} t_{\rm scl}$  & $12$ & $\mathrm{ns}$ \\ 
    Equilibrium film thickness & $h_* H$ & $1$ & $\mathrm{nm}$  \\  
  Mean filament thickness & $h_0 H$ & $5-15$ & $\mathrm{nm}$ \\
  Filament widths & $w H$ & $2-10$ & $\mathrm{nm}$ \\ 
  SiO$_2$ thickness & $H_{\rm s} H$ \cite{allaire_prf_2022} & $15$ & $\mathrm{nm}$  \\ 
  Room temperature & $T_{\rm a} T_{\rm melt}$ & $300$ & $\mathrm{K}$ \\  
    Activation Energy & $E$ \cite{metals_ref_book_2004} & $30.5$ & ${\rm kJ} \, {\rm mol}^{-1}$  \\ 
    Disjoining Pressure Exponents & $(n,m)$ & $(3,2)$ & 1  \\ \hline
 \end{tabular} }
 \caption{Parameters used for simulations based on ${\rm Cu}$.}
 \label{table:ref_paras}
\end{table}

\begin{table}[htb]
\centering
\setlength{\tabcolsep}{0.5em} % for the horizontal padding
{\renewcommand{\arraystretch}{1.17} % for the vertical padding
  \begin{tabular}{ | l | l | l | l |} \hline 
 \textbf{Dimensionless Numbers} & \textbf{Notation} & \textbf{Value} & \textbf{Expression}  \\ \hline
    Film Peclet Number & $\mbox{Pe}_{\rm f}$ & $1.29 \times 10^{-2}$ & $\left( \rho c \right)_{\rm f} U H/k_{\rm f}$ \\
    Substrate Peclet Number & $\mbox{Pe}_{\rm s}$ & $1.64$ & $(\rho c)_{\rm s} U H/k_{\rm s}$ \\
    % Substrate Peclet Number & $\mbox{Pe}_{\rm s}$ & $2.17 \times 10^{-2}$ & $(\rho c)_{\rm s} U \epsilon H/\kappa_{\rm s}$ \\
   Thermal Conductivity Ratio & $\mathcal{K}$ & $0.004$ & $k_{\rm s}/k_{\rm f}$ \\ 
   Dimensionless Activation Energy & $\mathcal{E}$ & $186.74$ & $E R / T_{\rm melt}$ \\
  Marangoni Number & ${\rm Ma}$ & $-0.36$ & $3\gamma_{\rm T} T_{\rm melt}/(2\gamma_{\rm f})$ \\ \hline
 \end{tabular} }
  \caption{Dimensionless parameters based on material parameters in Table~\ref{table:ref_paras}.}
 \label{table:nondim_paras_table}
\end{table}

\section{Domain size effect}
Here, we investigate the influence that the domain boundaries have on the results of the main text. It should be noted that the only heat loss mechanism of the system is through the lateral boundaries of the substrate, which are fixed at room temperature, $T_{\rm a}$. Therefore, it is important to consider the distance from the filament to the boundaries and its impact on the heating and subsequent evolution of the filaments.
\begin{figure}[H]
    \centering
    \includegraphics[width=0.6\textwidth]{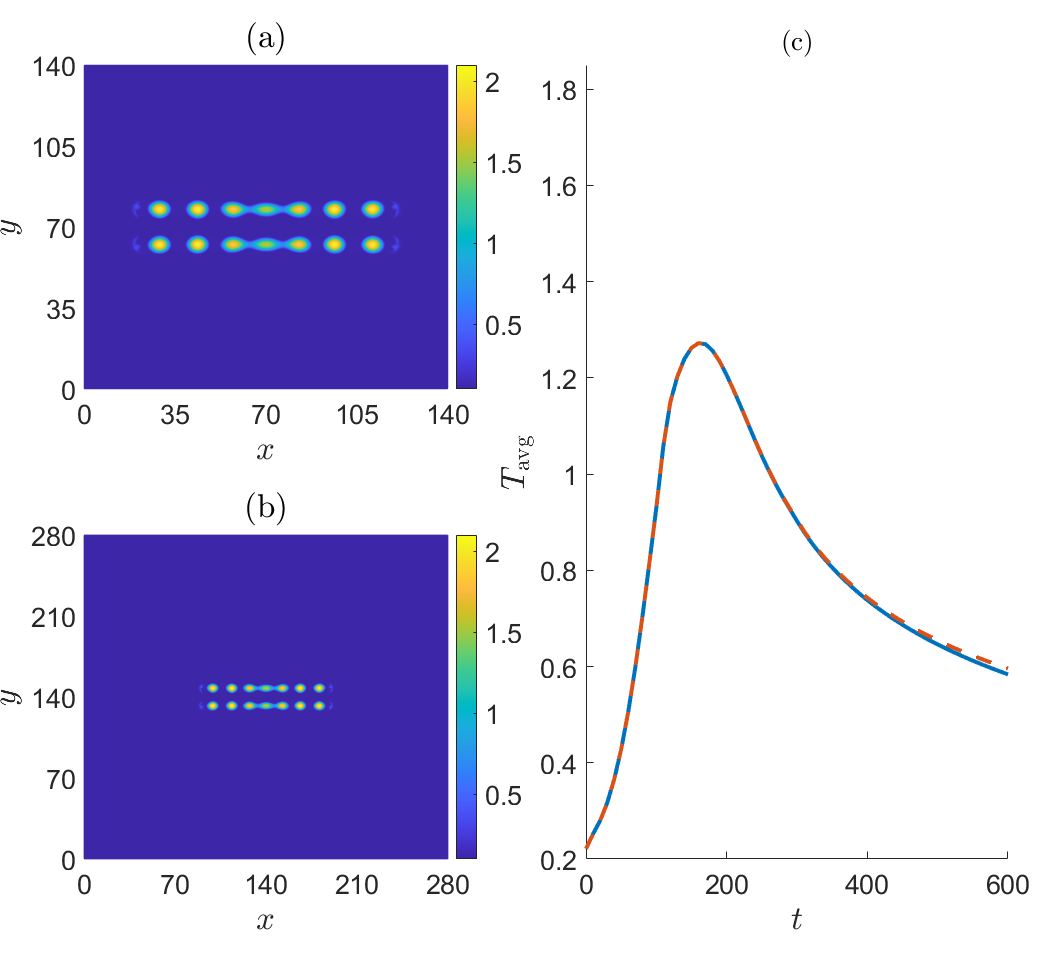}
    \caption{Influence of domain size.  Final configurations of two filaments for (a) the usual domain size, $P=140$ (reproduced from Fig.~2(b) in the main text), and (b) a larger domain, $P=280$. Part (c) shows the average filament temperatures of the configurations for (a) ({\color{blue} blue} solid line) and (b) ({\color{red} red} dashed line) along with the melting temperature (black solid line). See Animation 5 for fluid evolution and temperature fields.}
    \label{fig:domain_boundaries}
\end{figure}

Figure \ref{fig:domain_boundaries} shows the influence of domain size on the final configurations of the metals. Part (a) corresponds to the usual domain size ($P = 140$) (see Fig.~2(b) of the main text). In Part (b) the linear domain size is doubled (to 280) but the initial metal configuration is kept the same. Figure \ref{fig:domain_boundaries}(c) shows the average filament temperatures corresponding to the metals in Fig.~\ref{fig:domain_boundaries}(a) ({\color{blue} blue} solid line) and (b) ({\color{red} red} dashed line). Collectively (a-c), show that the domain size does not influence the heating and evolution of the filaments. Animation 5 reveals additionally that fluid and temperature evolution of both (a) and (b) are indistinguishable and since the simulations on smaller domains require less computational resources, we use such domains throughout. 

\section{Influence of Filament Volume}
\begin{figure}[t]
    \centering
    \includegraphics[width=0.85\textwidth]{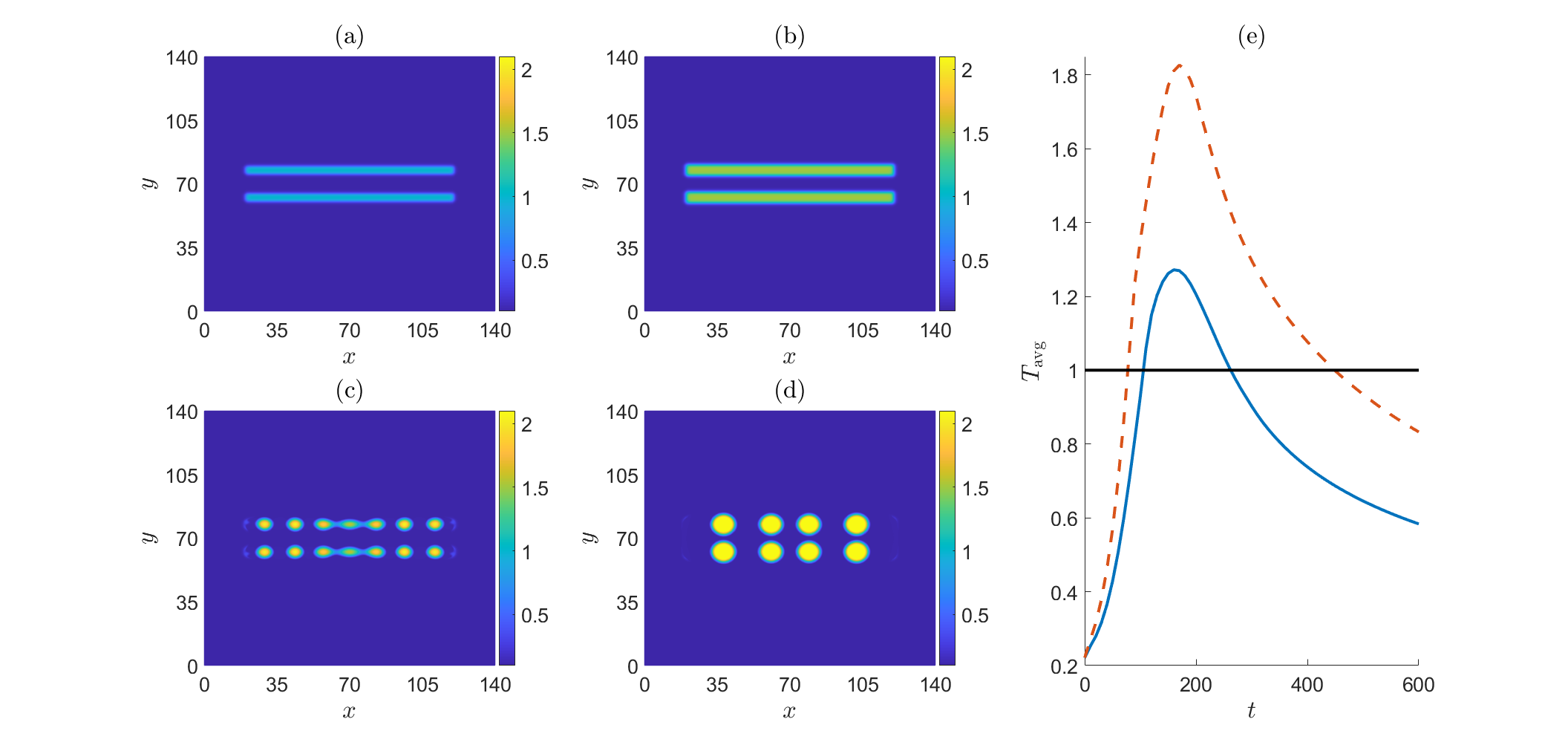}
    \caption{Influence of filament volume.  Initial filament configurations of heights (a) $h_0=1.0$ and (b) $h_0=1.5$ (both with the same height-to-width aspect ratio), and their respective final configurations (after melting, evolving, and solidifying) given in (c) and (d) (part (c) is reproduced from Fig.~2(b) in the main text). Part (e) shows the average filament temperatures for both $h_0=1.0$ ({\color{blue} blue} solid line) and $h_0=1.5$ ({\color{red} red} dashed line). The melting temperature is given by the black solid line. See also Animation 6 for fluid and temperature evolution. 
        }
    \label{fig:effect_of_volume}
\end{figure}

Here we investigate the influence that filament volume has on the heating and subsequent fluid evolution. Figures \ref{fig:effect_of_volume}(a-b) show the initial configuration of 2 filaments, separated by a distance $D=15$, with heights $h_0=1.0$ and $h_0=1.5$, respectively, but both (a) and (b) have constant height-to-width aspect ratio (i.e. the filament in (b) is wider to compensate for the increased height). Figure \ref{fig:effect_of_volume}(c) and (d) show the respective configurations once the filaments have melted, gone through fluid evolution and re-solidified. Part (e) shows the corresponding average filament temperatures for both (c) ({\color{blue} blue} solid line) and (d) ({\color{red} red} dashed line). We see from (c) and (d) that increased volume leads to higher temperatures and subsequently more dewetting with larger droplets. The filaments in (c), although melted, do not heat sufficiently to fully dewet, resulting only in the formation of partial droplets, which are much smaller than those in (d). Animation 6 reveals the increased dewetting speed of the larger volume filaments, due to the lower viscosity that results from the significantly increased temperature; complete dewetting of the larger filaments occurs prior to any breakup of the smaller filaments.

\section{Influence of Aspect Ratio}
\begin{figure}[t]
    \centering
    \includegraphics[width=\textwidth]{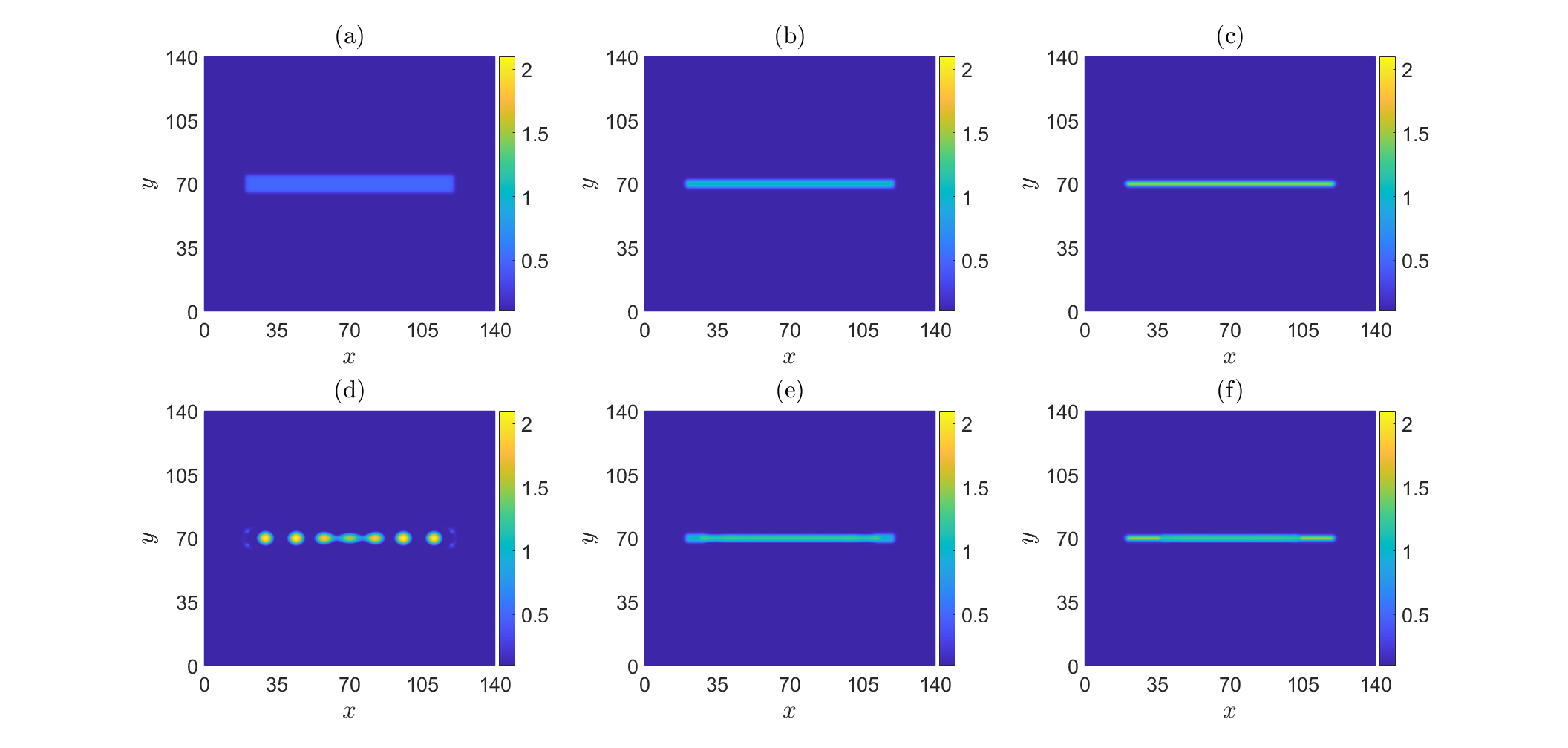}
    \caption{Influence of filament aspect ratio.  Initial configuration of single filaments for heights $h_0=0.5$ (a), $h_0=1.0$ (b), and $h_0=1.5$ (c), but with widths that keep volume constant (part (e) is reproduced from Fig.~2(a) in the main text). The final configurations (after re-solidification) are given in (d)-(f), respectively. See also Animation 7 for fluid and temperature evolution.}
    \label{fig:varying_initial_heights}
\end{figure}

Here we consider the effect that filament aspect ratio has on the fluid evolution. Figures \ref{fig:varying_initial_heights} (a-c) show an initial filament of heights $h_0=0.5$ (a), $h_0=1.0$ (b), and $h_0=1.5$ (c), with varying widths that conserve volume. Figures \ref{fig:varying_initial_heights}(d-f) show the same filaments after laser melting, fluid evolution, and solidification. We see from (d) that the thinner the filament is, the more it evolves. 

\begin{figure}[t]
    \centering
    \includegraphics[width=\textwidth]{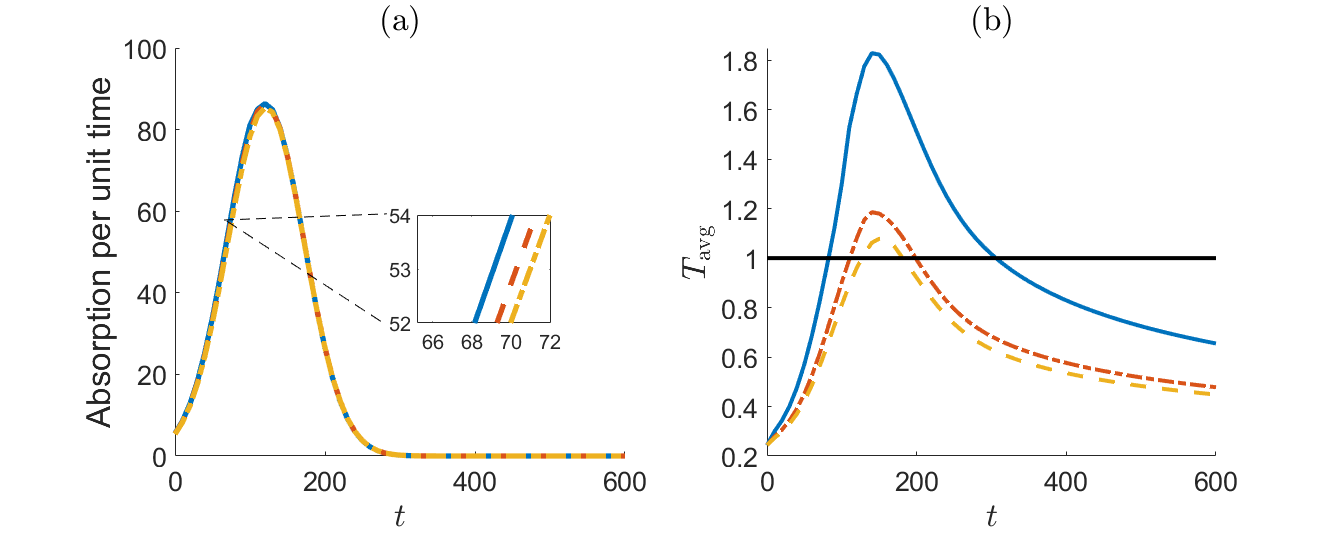}
    \caption{(a) Total energy absorbed by the fluid per unit time for $h_0=0.5$ ({\color{blue} blue} solid line), $h_0=1.0$ ({\color{red} red} dashed line) and $h_0=1.5$ ({\color{orange} orange} dashed line). Inset: Zoom-in of the absorption graph around melting time showing the differences between the curves. (b) Average filament temperature for the three cases in (a) along with the melting temperature (black solid line).}
    \label{fig:total_absorption}
\end{figure}

To clarify the effect that leads to faster evolution of thinner filaments, Fig.~\ref{fig:total_absorption}(a) shows the total energy from the laser absorbed by the metal calculated as
\begin{align*}
    \int\limits_0^P \int\limits_0^P h\overline{Q}dx dy=\int\limits_0^P \int\limits_0^P A(h) F(t)(1-R(h))(1-e^{-\alpha_{\rm f} h}) dxdy,
\end{align*}
and Fig. \ref{fig:total_absorption}(b) shows the corresponding average filament temperatures from the simulations in Fig.~\ref{fig:varying_initial_heights}. From Fig.~\ref{fig:total_absorption}(a) we see that the filament shown in Fig.~\ref{fig:varying_initial_heights}(a) absorbs slightly more energy than the others, and is therefore the hottest (see Fig.~\ref{fig:total_absorption}(b)), and the furthest evolved, see Fig.~\ref{fig:varying_initial_heights}(d). We note that this scenario is not obvious since the energy absorption is thickness-dependent (thicker films absorb more energy).  For filament geometry, however, there is a competing effect of larger surface area, leading to larger total absorption for the wider filament.  For the particular set of parameters considered, our results show that the larger initial surface area outweighs the decreased absorption (per unit area), and therefore, the thinner filament is slightly hotter and faster evolving. The fluid and temperature evolution, seen in Animation 7, reveals this increased evolution of the thinner filament, despite the additional time needed for it to retract in the transverse ($y$) direction.

\section{Multiple filaments}

\begin{figure}[t]    
\centering
    \includegraphics[width=\textwidth]{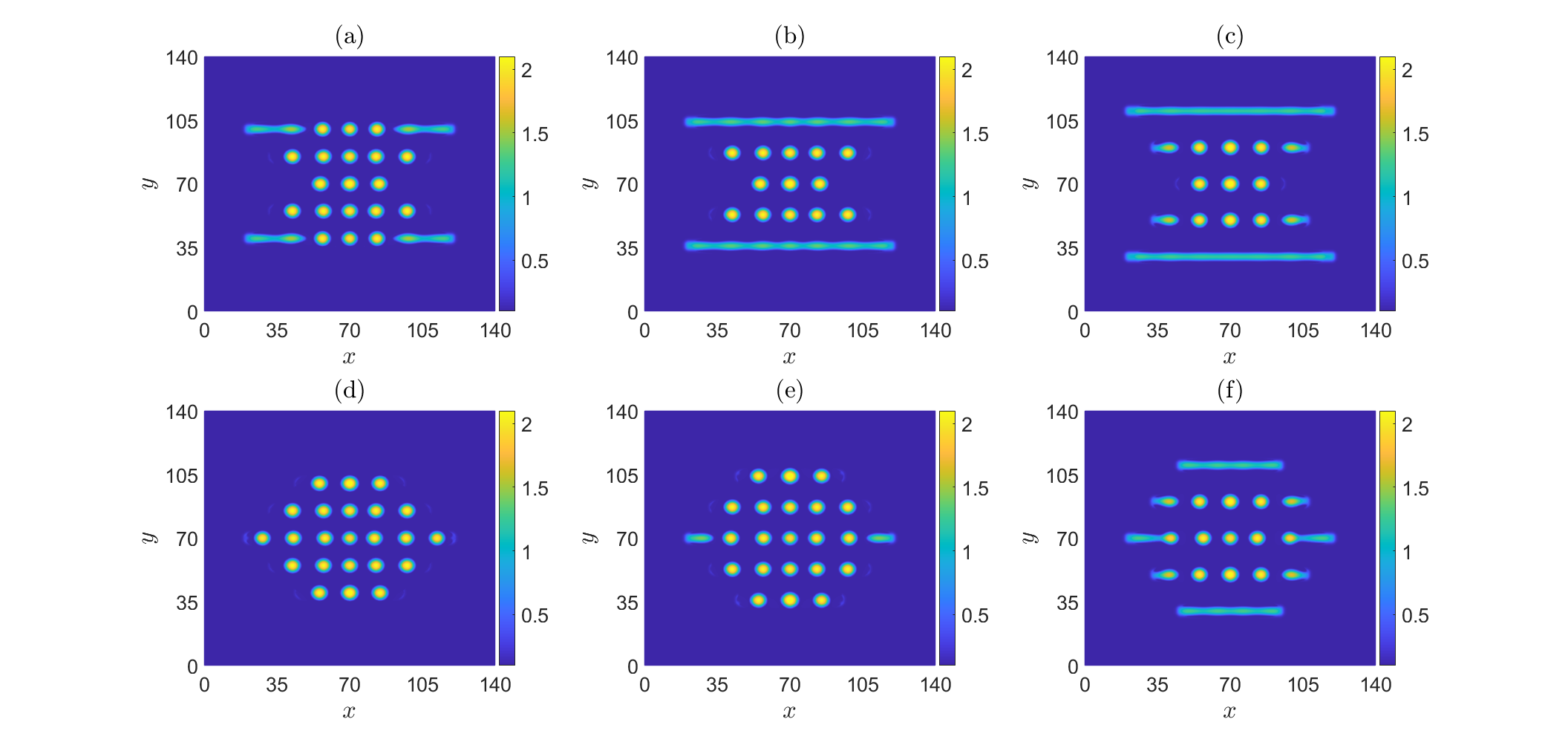}
    \caption{Influence of filament length and placement. (a-c) Final metal configurations for a short filament of length $L=50$ surrounded by sequentially longer filaments, $L=75$ and $L=100$, all at distances (a) $D=15$, (b) $D=17$, and (c) $D=20$ apart. (d-f) Final metal configurations for a long filament, $L= 100$, surrounded by sequentially shorter filaments, $L=75$ and $L=50$, all at distances (d) $D=15$, (e) $D=17$, and (f) $D=20$ apart. See also Animations 8 and 9 for fluid and temperature evolution.
        }
    \label{fig:asymmetries}
\end{figure}

Figure \ref{fig:asymmetries} illustrates that filament instability can be initiated asymmetrically. Here, we place five filaments at distances $D=15, 17, 20$ with longer filaments either on the exterior (a-c) or the interior (d-f). From parts (a-c) we see a similar outcome as in Fig.~3 of the main text, where increasing the interfilament distance leads to slower evolution of the exterior filaments. 

Surrounding a long filament by shorter ones leads to a different outcome. Although we see nearly full dewetting in (d) and (e), we obtain localized dewetting of some of the filaments in (f), see also Animations 8 and 9. For the case (f), we observe that the outside filaments are frozen in place. The second filaments from the boundaries along the $y-$axis are almost fully dewetted, although the final configuration shows that some filament remains near the ends. The central filament shows a mixture of dewetting that occurs primarily due to the disparity between the length of the central filament and its neighbors. Animations 8 and 9 show that the nature of instability development is different, compared to what has been seen so far: for the central filament, the instability is not of pearling type (propagating from the filament ends), but the filament actually breaks up along its length. The reader is directed to~\cite{dk_pof07} for detailed discussion of the similarities and differences between the two instability types. It should be noted (see the Anomations 8 and 9) that the formation of a droplet in the very center of Fig.~\ref{fig:asymmetries}(f) is nontrivial due to the competition between the edge retraction of the inner-most filament and its cooling. 

We note that the average temperatures of the filaments do not, in this case, provide any additional insight into the behavior in Fig.~\ref{fig:asymmetries}. Instead, animations of temperature fields turn out to be more useful for the purpose of understanding the filament evolution. Animation 9 reveals the nontrivial competition between the cooling and breakup seen in, for example, Fig.~\ref{fig:asymmetries}(f). The interiors of the inner-most filaments in (f) become much hotter than the edges, and therefore solidify much later than the edges, which only begin dewetting. 
This indicates a competition between the liquid lifetime and the time scale of metal edge retraction which all vary spatially because of the spatial variation in temperature and viscosity. 

We note that other geometries could be considered; however, the geometry considered in Fig.~\ref{fig:asymmetries} already shows that placement of additional metal geometries can provide precise control over instability development. 

\bibliography{films}